\documentclass[twoside,twocolumn,9pt]{article}
\usepackage{extsizes}
\usepackage[super,sort&compress,comma]{natbib} 
\usepackage[version=3]{mhchem}
\usepackage[left=1.5cm, right=1.5cm, top=1.785cm, bottom=2.0cm]{geometry}
\usepackage{balance}
\usepackage{mathptmx}
\usepackage{sectsty}
\usepackage{graphicx} 
\usepackage{lastpage}
\usepackage[format=plain,justification=justified,singlelinecheck=false,font={stretch=1.125,small,sf},labelfont=bf,labelsep=space]{caption}
\usepackage{float}
\usepackage{fancyhdr}
\usepackage{fnpos}
\usepackage[english]{babel}
\addto{\captionsenglish}{%
  
}
\usepackage{array}
\usepackage{droidsans}
\usepackage{charter}
\usepackage[T1]{fontenc}
\usepackage[usenames,dvipsnames]{xcolor}
\usepackage{setspace}
\usepackage[compact]{titlesec}
\PassOptionsToPackage{hyphens}{url}\usepackage{hyperref}

\definecolor{cerclegreen}{HTML}{139F46}
\definecolor{DarkTurquoise}{HTML}{00CED1}

\usepackage{epstopdf}

\definecolor{cream}{RGB}{222,217,201}

\begin{document}

\pagestyle{fancy}
\thispagestyle{plain}
\fancypagestyle{plain}{
\renewcommand{\headrulewidth}{0pt}
}

\makeFNbottom
\makeatletter
\renewcommand\LARGE{\@setfontsize\LARGE{15pt}{17}}
\renewcommand\Large{\@setfontsize\Large{12pt}{14}}
\renewcommand\large{\@setfontsize\large{10pt}{12}}
\renewcommand\footnotesize{\@setfontsize\footnotesize{7pt}{10}}
\makeatother

\renewcommand{\thefootnote}{\fnsymbol{footnote}}
\renewcommand\footnoterule{\vspace*{1pt}%
\color{cream}\hrule width 3.5in height 0.4pt \color{black}\vspace*{5pt}} 
\setcounter{secnumdepth}{5}

\makeatletter 
\renewcommand\@biblabel[1]{#1}            
\renewcommand\@makefntext[1]%
{\noindent\makebox[0pt][r]{\@thefnmark\,}#1}
\makeatother 
\renewcommand{\figurename}{\small{Fig.}~}
\sectionfont{\sffamily\Large}
\subsectionfont{\normalsize}
\subsubsectionfont{\bf}
\setstretch{1.125} 
\setlength{\skip\footins}{0.8cm}
\setlength{\footnotesep}{0.25cm}
\setlength{\jot}{10pt}
\titlespacing*{\section}{0pt}{4pt}{4pt}
\titlespacing*{\subsection}{0pt}{15pt}{1pt}

\fancyfoot{}
\fancyfoot[LO,RE]{\vspace{-7.1pt}\includegraphics[height=9pt]{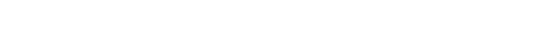}}
\fancyfoot[CO]{\vspace{-7.1pt}\hspace{11.9cm}\includegraphics{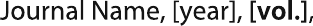}}
\fancyfoot[CE]{\vspace{-7.2pt}\hspace{-13.2cm}\includegraphics{head_foot/RF}}
\fancyfoot[RO]{\footnotesize{\sffamily{1--\pageref{LastPage} ~\textbar  \hspace{2pt}\thepage}}}
\fancyfoot[LE]{\footnotesize{\sffamily{\thepage~\textbar\hspace{4.65cm} 1--\pageref{LastPage}}}}
\fancyhead{}
\renewcommand{\headrulewidth}{0pt} 
\renewcommand{\footrulewidth}{0pt}
\setlength{\arrayrulewidth}{1pt}
\setlength{\columnsep}{6.5mm}
\setlength\bibsep{1pt}

\makeatletter 
\newlength{\figrulesep} 
\setlength{\figrulesep}{0.5\textfloatsep} 

\newcommand{\topfigrule}{\vspace*{-1pt}%
\noindent{\color{cream}\rule[-\figrulesep]{\columnwidth}{1.5pt}} }

\newcommand{\botfigrule}{\vspace*{-2pt}%
\noindent{\color{cream}\rule[\figrulesep]{\columnwidth}{1.5pt}} }

\newcommand{\dblfigrule}{\vspace*{-1pt}%
\noindent{\color{cream}\rule[-\figrulesep]{\textwidth}{1.5pt}} }

\makeatother

\twocolumn[
  \begin{@twocolumnfalse}
{
\includegraphics[width=18.5cm]{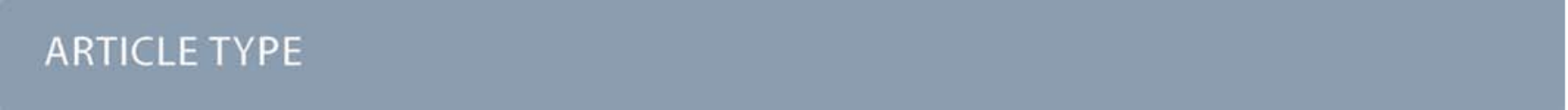}}\par
\vspace{1em}
\sffamily
\begin{tabular}{m{4.5cm} p{13.5cm} }

\includegraphics{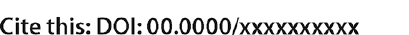} & \noindent\LARGE{\textbf{Estimating the heating of complex nanoparticle aggregates for magnetic hyperthermia}} \\

\vspace{0.3cm} & \vspace{0.3cm} \\

 & \noindent\large{Javier Ortega-Julia,\textit{$^{a, b}$} Daniel Ortega\textit{$^{b, c, d}$} and Jonathan Leliaert $^{\ast}$\textit{$^{a}$}} \\

\includegraphics{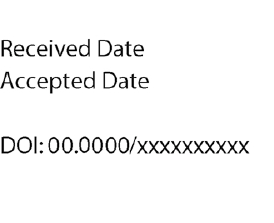} & \noindent\normalsize{Understanding and predicting the heat released by magnetic nanoparticles is central to magnetic hyperthermia treatment planning. These nanoparticles tend to form aggregates when injected in living tissues, which alters their response to the applied alternating magnetic field and prevents predicting the released heat accurately. We performed an \textit{in silico} analysis to investigate the heat released by nanoparticle aggregates featuring different size and fractal geometry factors. By digitally mirroring aggregates seen in biological tissues, we found that the average heat released per particle stabilizes starting from moderately small aggregates, facilitating the estimates for their larger counterparts. Additionally, we studied the heating performance of particle aggregates over a wide range of fractal parameters. We compared this result with the heat released by non-interacting nanoparticles to quantify the reduction of heating power after being instilled into tissues. This set of results can be used to estimate the expected heating \textit{in vivo} based on the experimentally determined nanoparticle properties.
} \\

\end{tabular}

 \end{@twocolumnfalse} \vspace{0.6cm}
]


\renewcommand*\rmdefault{bch}\normalfont\upshape
\rmfamily
\section*{}
\vspace{-1cm}

\footnotetext{\textit{$^{*}$~E-mail: jonathan.leliaert@ugent.be}}
\footnotetext{\textit{$^{a}$~Department of Solid State Sciences, Ghent University, Ghent, Belgium. }}
\footnotetext{\textit{$^{b}$~IMDEA Nanoscience, Faraday 9, 28049 Madrid, Spain. }}
\footnotetext{\textit{$^{c}$~Condensed Matter Physics Department, Faculty of Sciences, Campus Universitario Río San Pedro s/n. 11510 Puerto Real, Cádiz, Spain. }}
\footnotetext{\textit{$^{d}$~Institute of Research and Innovation in Biomedical Sciences of the Province of Cádiz (INiBICA), University of Cádiz, 11009 Cádiz, Spain. }}




\section{Introduction}
Magnetic nanoparticles dissipate heat upon the application of an alternating magnetic field (AMF). This localized heat can be harnessed in cancer therapy as an adjuvant to other first line treatments, giving rise to what is known as magnetic particle hyperthermia, or simply magnetic hyperthermia \cite{Ortega2013, espinosa2016duality}. Over the last years, the main components of this therapy - namely biocompatible magnetic nanoparticles and AMF applicators - have experienced notable advances \cite{Cabrera2019}, contributing to improve its nominal effectiveness and increase the likelihood of a wider clinical adoption. Thanks to these advances, and together with the specification of new ISO standards\cite{ISO1}, the use of magnetic hyperthermia is spreading worldwide, which in turn contributes towards its clinical implementation \cite{Rubia2021a}.
Given the strict requirements to approve medical devices, drugs and combination products \cite{regu1,regu2}, the control over the hyperthermia process and, consequentially, the safety of the procedure is still one of the most critical aspects to improve \cite{Rubia2021b,Herrero2022,Felista2021}.
In order to ensure a safe and efficient treatment, the goal is to achieve the therapeutic temperature in the tumor area - typically the dogmatic 43 ºC - to induce apoptosis to malignant cells without damaging the healthy surrounding tissue. 
Therefore, understanding the physical properties of the nanoparticles and their response to the applied AMF, mainly the amount of generated heat, is a critical step in determining the best possible treatment for a particular tumor. In this field, some seminal \textit{in silico} analyses were performed to simulate \textit{in vivo} nanoparticle heating exchange on a millimeter length scale\cite{Gellermann2000,Sreenivasa2003}. However, these methods are lacking single particle information, and therefore require additional input data to correctly model the generated heat at each location. Providing accurate estimates of this heat is hampered by the fact that nanoparticles form complex aggregates after cell uptake in biological tissues due to the interactions with proteins \cite{Aggarwal2009}. The local heating achieved at the tumor site is not only determined by the local average iron concentration\cite{Pankhurst2018}, but also by the intratumoral three-dimensional distribution of nanoparticles within agregates, which has been the subject of numerous \textit{in vivo} studies\cite{Zadnik2014,Dahring2015,Tansi2021}. The inherent difficulty in determining precisely the structure and distribution of these aggregates has led to the search for indirect methods of monitoring them after injection of the nanoparticles - mainly intratumourally -, such as the use of infrared thermal imaging\cite{Kuboyabu2016} or magnetic particle imaging (MPI)\cite{perspective}. There is only limited understanding of single particle heating within such aggregates, which is especially problematic because it is known that this aggregation greatly affects the released heat. As shown by \citet{Etheridge2014} and \citet{Dahring2015}, there are significant differences in heating for various clusters in \textit{in vivo} experiments as compared to \textit{in vitro} measurements. Therefore, using \textit{in silico} analysis to study the amount and location of the generated heat within aggregates can lead to a better understanding of the heating phenomena at the macroscale.
The work done by \citet{carrey2011}, \citet{Raikher2014} and \citet{Munoz2015} show several micromagnetic and Monte Carlo models to estimate and optimize single particle heating. However, because the nanoparticle heating behavior in aggregates is strongly affected by interparticle interactions, these models are not directly applicable to estimate the heat generated by nanoparticle clusters. Based on the work by \citet{munoz2020}, \citet{Leliaert2021} have developed the necessary methodology to address this challenge, allowing to calculate the heat dissipated by individual, interacting, particles due to field and thermally induced switching. Building upon this new methodology, the present work aims to close the gap between single- and multiple-particle simulations by studying the average heating per nanoparticle in clusters. We show that it is possible to infer the heating released by large aggregates of nanoparticles from that of their smaller counterparts.

\section{Methods}
The magnetization dynamics, and consequently the heat released by the magnetic nanoparticles, will be studied within the micromagnetic framework.
The nanoparticles used in magnetic hyperthermia typically have a size of a few tens of nanometers \cite{Pankhurst2003}, which is comparable to the exchange length for materials commonly considered (like ferrimagnetic iron oxides). The particles therefore are uniformly magnetized\cite{Tong2019}, i.e. a single domain state, and can be approximated by a single macrospin. In micromagnetism, the magnetization dynamics are described by the Landau-Lifshitz-Gilbert equation, Eq. (\ref{eq:LLG}):
\begin{equation}
  \frac{d\mathbf{m}}{dt}= -\frac{\gamma }{1+\alpha ^{2}} [\mathbf{m} \times \mathbf{B}_\mathrm{eff}+\alpha \mathbf{m}\times(\mathbf{m} \times \mathbf{B}_\mathrm{eff})]
  \label{eq:LLG}
\end{equation}
In this equation, $\gamma$ denotes the gyromagnetic ratio and $\alpha$ the Gilbert damping constant. It describes the time evolution of the magnetization $\mathbf{m}$ as a function of the effective field $\mathbf{B}_\mathrm{eff}$, comprising an externally applied field; the anisotropy field {\(\mathbf{B}_\mathrm{anis}=\frac{2K}{M_\mathrm{s}}(\mathbf{m}\cdot \mathbf{u})\mathbf{u})\)}
where $K$ and $\mathbf{u}$ denote the strength and direction of the uniaxial anisotropy, respectively, and $M_\mathrm{s}$ the saturation magnetization; the dipolar field $\mathbf{B}_\mathrm{int}$ which accounts for the interparticle interactions; and the stochastic thermal field $\mathbf{B}_\mathrm{th}$ which account for the effects of thermal fluctuations at nonzero temperatures\cite{leliaert2017,lyberatos1993,brown1963thermal}.

All simulations are performed using the macrospin simulation tool Vinamax \cite{Vinamax}, and the heat $\mathcal{E}$ released by each of the particles was determined by integrating the following equation\cite{Leliaert2021}:
\begin{equation}
    \frac{d\mathcal{E}}{dt}=\frac{\alpha\gamma M_s}{1+\alpha^2}\left(\mathbf{m}\times\mathbf{B}_\mathrm{eff}\right)^2-M_s\mathbf{B}_\mathrm{th}\cdot \frac{d\mathbf{m}}{dt}
    \label{eq:heating}
\end{equation}

In biological tissues, nanoparticles tend to form aggregates that can be accurately described as fractals \cite{Cai2017,Prasher2006,hovorka2017thermal}, characterized by the fractal equation \cite{Forrest1979,filippov2000fractal}, Eq. (\ref{eq:fractal}):
\begin{equation}
  n_{p}=  k_\mathrm{f} \left(\frac{R_\mathrm{g}}{r_\mathrm{p}}\right)^{D_\mathrm{f}}
  \label{eq:fractal}
\end{equation}
where $n_\mathrm{p}$ and $r_\mathrm{p}$ represents the number of particles in the cluster and their radius, $R_\mathrm{g}$ is the radius of gyration and $D_\mathrm{f}$ and $k_\mathrm{f}$ are the fractal dimension and fractal prefactor, respectively. The fractal dimension contains most of the information on the cluster shape and ranges between 1 to 3, with 1 corresponding to a straight line and 3 to a perfectly spherical cluster. Most fractals found in biological tissue have a dimension ranging from 1.6 to 2.5 \cite{Hogan2016,Etheridge2014}. The physical meaning\cite{cai1995,Bau2010} of the fractal prefactor $k_\mathrm{f}$ and its numerical value are still a topic under discussion. For instance, \citet{Wozniak2012} references values from 1.23 to 3.5, while \citet{Hogan2016} and \citet{Etheridge2014} report values from 1.2 to 2, and \citet{Pratsinis2012} even suggest values as low as 0.41. In this work we will consider values between 0.5 and 1.9.
To generate the fractal geometry with the desired number of particles, size and fractal dimension, we used FracVal\cite{Moran2019}, an algorithm for cluster generation. This tool allowed us to generate aggregates for $D_\mathrm{f}$ between 1 and 3, which were subsequently used as input for the micromagnetic simulations.

To study how the generated heat depends on the number of particles in the aggregate, we simulated different fractals with $n_\mathrm{p}$ ranging between 1 and 100. In particular, we considered three different sets of fractal parameters using the values reported by \citet{Etheridge2014}, which are based on measurements of real particle aggregates in cells:
$D_\mathrm{f}$=1.8, $k_\mathrm{f}$=1.3; $D_\mathrm{f}$=2, $k_\mathrm{f}$=1.5 and $D_\mathrm{f}$=2.1,  $k_\mathrm{f}$=1.3.

Each of these sets of parameters was used to generate a hundred different clusters, thereby accounting for the different orientations and configurations the particles can present, and which may affect the generated heat. The simulations were performed using monodisperse spherical particles with  a diameter equal to 22 nm. The material parameters are those of maghemite \cite{Coey2001}: $\alpha= 0.5$, $M_\mathrm{s}= 446$ $ \mathrm{kA m^{-1}}$, $K_\mathrm{anis}=5$ $\mathrm{kJ m^{-3}}$, with the uniaxial anisotropy axis randomly oriented in each particle. We performed simulations at temperatures of both 0 K and 300 K, using a sinusoidally varying field with intensities of 5, 10, 15, 20, 25 and 30 mT, and a frequency of 300 kHz. 

These simulations allowed us to define a cluster size - containing about 20 particles - above which the average heat per particle stagnates. 

Subsequently, we investigated how the heating is affected by the fractal parameters by expanding the number of considered $D_\mathrm{f}$ and $k_\mathrm{f}$. A total to 31 different geometries, whose parameters can be seen in Figure \ref{fig:HeatVsPara.png}, were studied using the same physical properties mentioned above under the application of a 25 mT field. In this study, we used clusters comprising 50 particles, for which the heat per particle is independent of the number of particles.

\section{Results and discussion}
\subsection{Average heat per particle as a function of cluster size}
We investigated the cluster heating as a function of the cluster size at both 0K (the athermal case) and 300 K, thereby including thermal switching, similar to what happens in any real world application. The athermal case is also included in our analysis because it gives additional insight into the energy barriers and allows to gain a deeper understanding of what happens at nonzero temperatures.
These results are presented in Figure \ref{fig:Heat0Kwitherrors.pdf}, (0 K), and Figure \ref{fig:Heat300Kwitherrors.pdf} (300 K). Panels (a-f) show the results obtained for different field intensities while the clusters (shown at the top of the figure) were generated using the different (color-coded) sets of fractal parameters. 

\subsubsection{Athermal case}
\label{sec:0K}

\begin{figure*}[ht]
\centering
  \includegraphics[width=17.2cm]{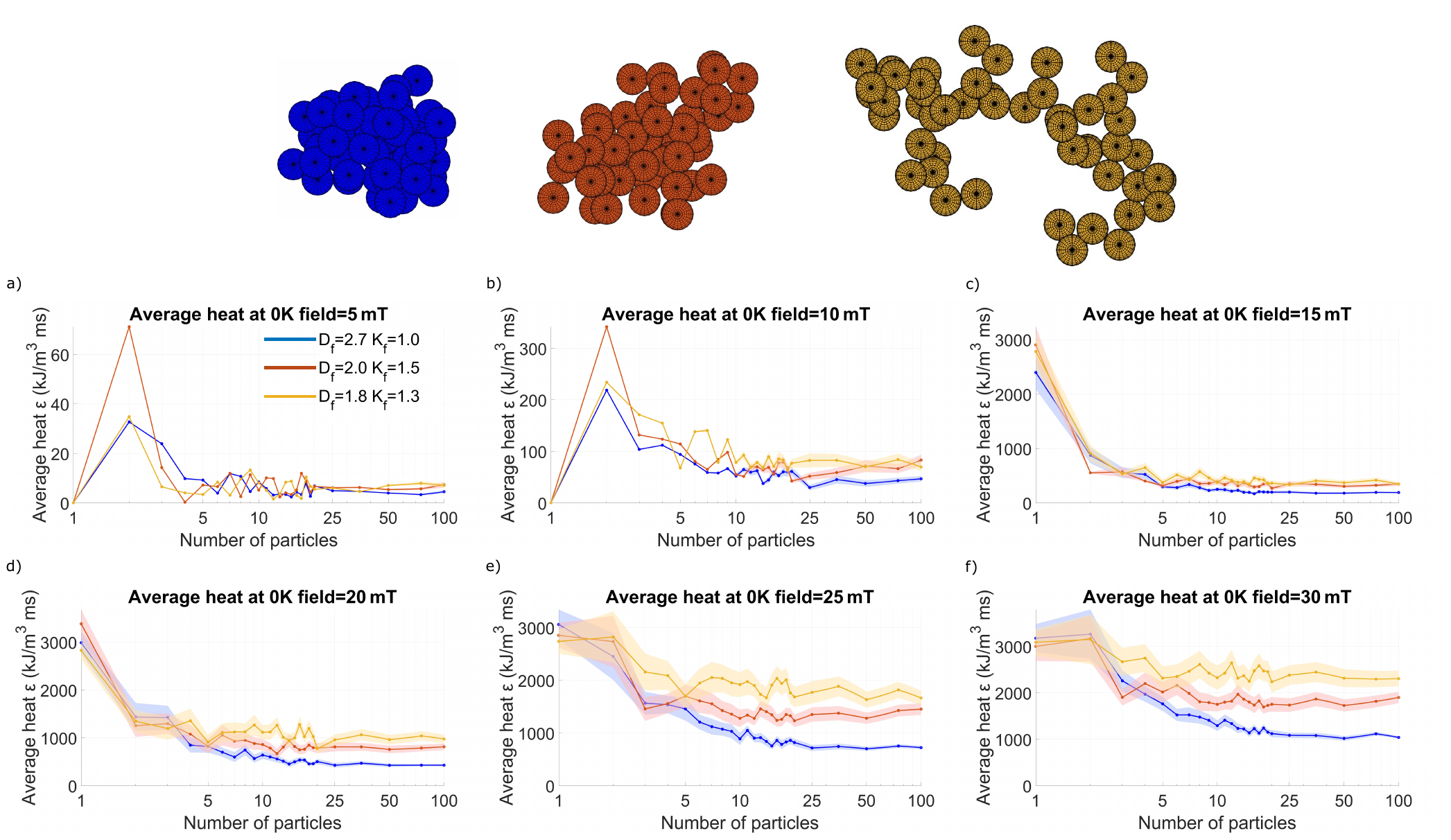}
  \caption{
  Average generated heat per particle, and the uncertainty on this value (shaded regions) released by the different color-coded clusters (colored lines) as a function of the number of particles at 0 K in the presence of a sinusoidally varying applied field with a frequency of 300 kHz and  amplitudes ranging from 5 mT to 30 mT. Three different sets of fractal parameters are shown: $D_\mathrm{f}=2.7$ $k_\mathrm{f}=1$ (blue); $D_\mathrm{f}=2$ $k_\mathrm{f}=1.5$ (red) and $D_\mathrm{f}=1.8$ $k_\mathrm{f}=1.3$ (yellow). }
  \label{fig:Heat0Kwitherrors.pdf}
\end{figure*}

 We can see in Fig. \ref{fig:Heat0Kwitherrors.pdf} that for the lower field intensities of 5 mT and 10 mT [panels (a) and  (b)], the particles release very little heat as the field generally is not sufficiently strong to help the magnetization overcome the anisotropy barriers and induce magnetic switching.
As a consequence, single particle "clusters" do not release any noticeable heat. In contrast, for clusters containing more than one particle, interparticle interactions do result in some heating, with a peak for 2-particle clusters, followed by a gradual decline for larger clusters. It is deceptive to consider this peak as a proper increase in the heating (as compared to the single particles, which do not display any heating at all) because out of all the possible anisotropy configurations in 2-particle clusters, only very few result in an energy landscape with sufficient reduction of the switching energy barrier 
to produce some heat. 
Note that for 2-particle clusters the three considered cluster geometries are identical (a 2-particle chain) and the seemingly larger peak for the cluster with $D_\mathrm{f}=2.0$ (in red) as compared to the others is an artefact originating in the very low amount of produced heat.  

Once the field is sufficiently large, at 15, 20, 25 and 30 mT, [Panels (c), (d), (e) and (f)], the particles start releasing heat.  
The heat released by single particle clusters equals about 3000 kJ m$^{-3}$ ms$^{-1}$ \footnote{A note on the units used: we report the particle heating in units of kJ m$^{-3}$ms$^{-1}$, meaning that this is the amount of energy (in kilojoules) that gets dissipated as heat per cubic meter of \emph{magnetic material} per millisecond. This heating metric in itself gives no indication of the temperature reached as it  gets determined by the heat exchange with the surrounding cells or tissues\cite{Rabin2002}. However, it does give a good estimate of the amount of thermal energy that gets added at a certain location to the body, from which the temperature increase can be estimated in a next step.}, which is consistent with the value expected from the Stoner-Wohlfarth model (see appendix) as about 1/4th of the value given by $4 M_s B_c = 40$ kJ m$^{-3}$ with $B_c=\frac{2K}{M_s}$. Using $f=300$ kHz, this means that a heat of about 3000 kJ m$^{-3}$ ms$^{-1}$ is produced.

The three differently colored lines correspond to clusters with different fractal parameters, displaying different heating. However, before looking into these differences, we remark that these figures display a few common trends: at 15 mT and 20 mT, the 2-particle clusters heat less than the single particles, whereas at 25 mT and 30 mT the heating is more or less equal to the one particle case. 
In all cases, the heat generation then goes down as function of the number of particles and reaches a more or less constant value for clusters containing about 20 particles.
This trend can be explained as follows: for most 2-particle configurations, the magnetostatic interaction causes the switching energy barrier to be larger than for the single particle case, therefore resulting in less heating at low fields. However, when these barriers can be overcome at larger fields, a larger amount of heat can be generated as compared to that from single particles. For larger clusters, containing as few as 3 particles, the interactions result in a more isotropic contribution to the magnetic energy landscape, therefore progressively flattening out these large energy barriers and resulting in a decline in the heating. 

\begin{figure}[ht]
\centering
  \includegraphics[width=8.6cm]{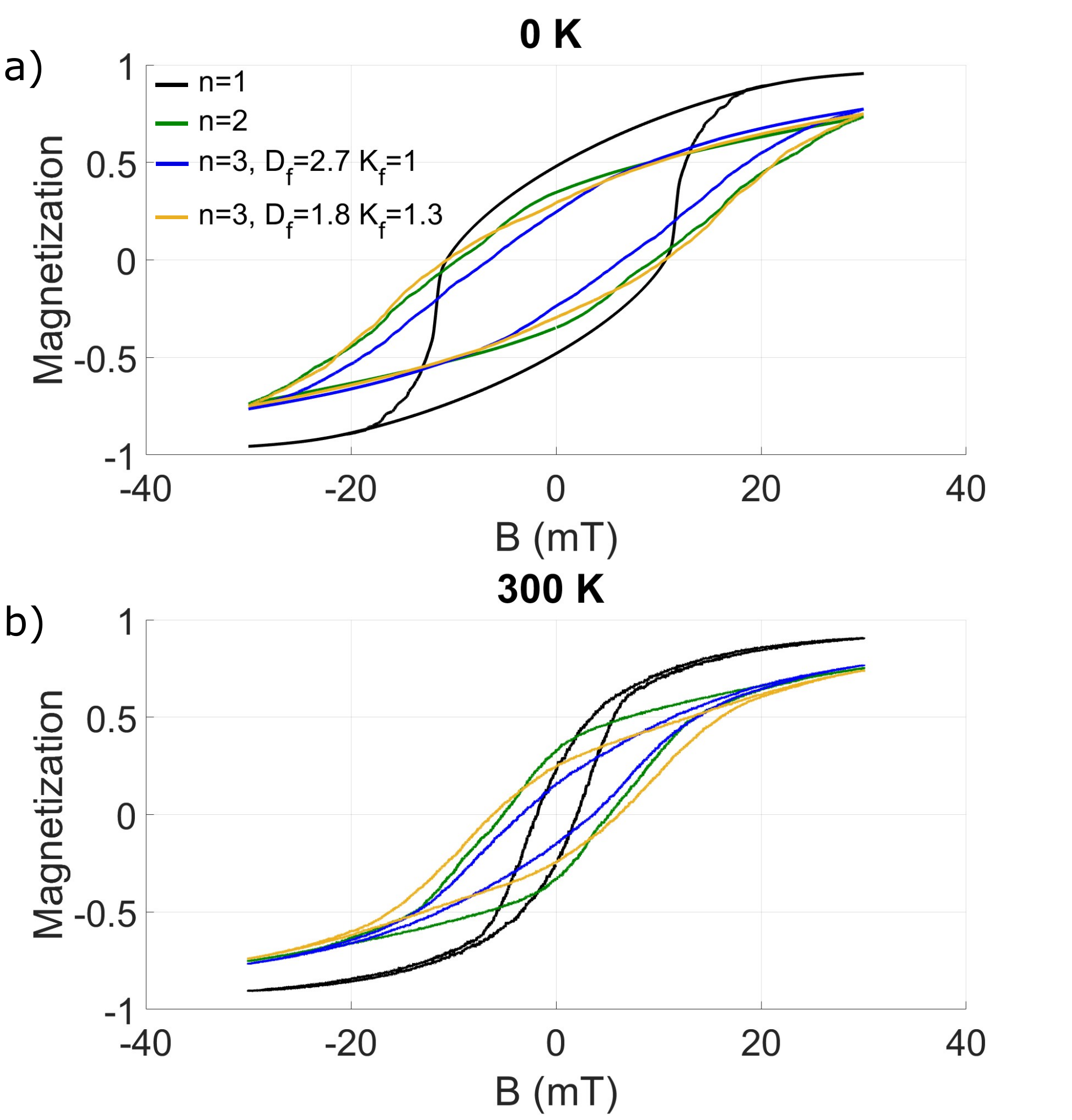}
  \caption{Hysteresis loops obtained from 1, 2 and 3 particles (the latter with fractal parameters $D_f$=2.7 $k_f$=1 and $D_f$=1.8 $k_f$=1.3) at 30 mT field intensity and at 0K (top) and 300K (bottom)}
  \label{HystLoops.pdf}
\end{figure}

This interpretation of the data is confirmed by the hysteresis loop shown in Fig. \ref{HystLoops.pdf} a). For single particles, the hysteresis loop already closes slightly below 20 mT, and larger fields will not result in any additional heating. The other loops do not display this behavior, because irreversible switches are still happening at these larger fields, indicating the presence of higher energy barriers.

A cluster size of 20 particles is already sufficient for each of the particles to feel the same local environment due to the interactions with their neighbours, such that the average heat released per particle reaches a constant value. This is one of the main conclusions of this work, as it allows to estimate the heating of very large clusters based on a modestly sized simulation, provided that the fractal dimensions of the cluster under study are known.

Indeed, the heat released presents important variations depending on the fractal dimensions. Specifically, the compact cluster represented by a blue line releases less heat than the most elongated one, represented by the yellow line. This is readily understood because the particles in the compact, spherical, clusters are influenced by neighbours in all directions, which leads to an isotropic energy landscape and consequently a decrease of the switching energy barriers, and less heating. The elongated, chain-like, clusters, on the other hand, mostly have neighbours in one direction, such that their interactions give rise to a shape anisotropy which conserves the energy barriers and thus results in more heating than their compact counterparts. These conclusions are also supported by the hysteresis loops shown in Fig. \ref{HystLoops.pdf} a), in which the narrowest loop corresponds to the most compact cluster geometry and the widest loop (except for the single particle case) corresponds to the most elongated clusters.
The influence of the fractal parameters on the heating of the clusters is further investigated in Section \ref{sec:parameters}.

\subsubsection{Nonzero temperatures}
\begin{figure*}[ht]
\centering
  \includegraphics[width=17.2cm]{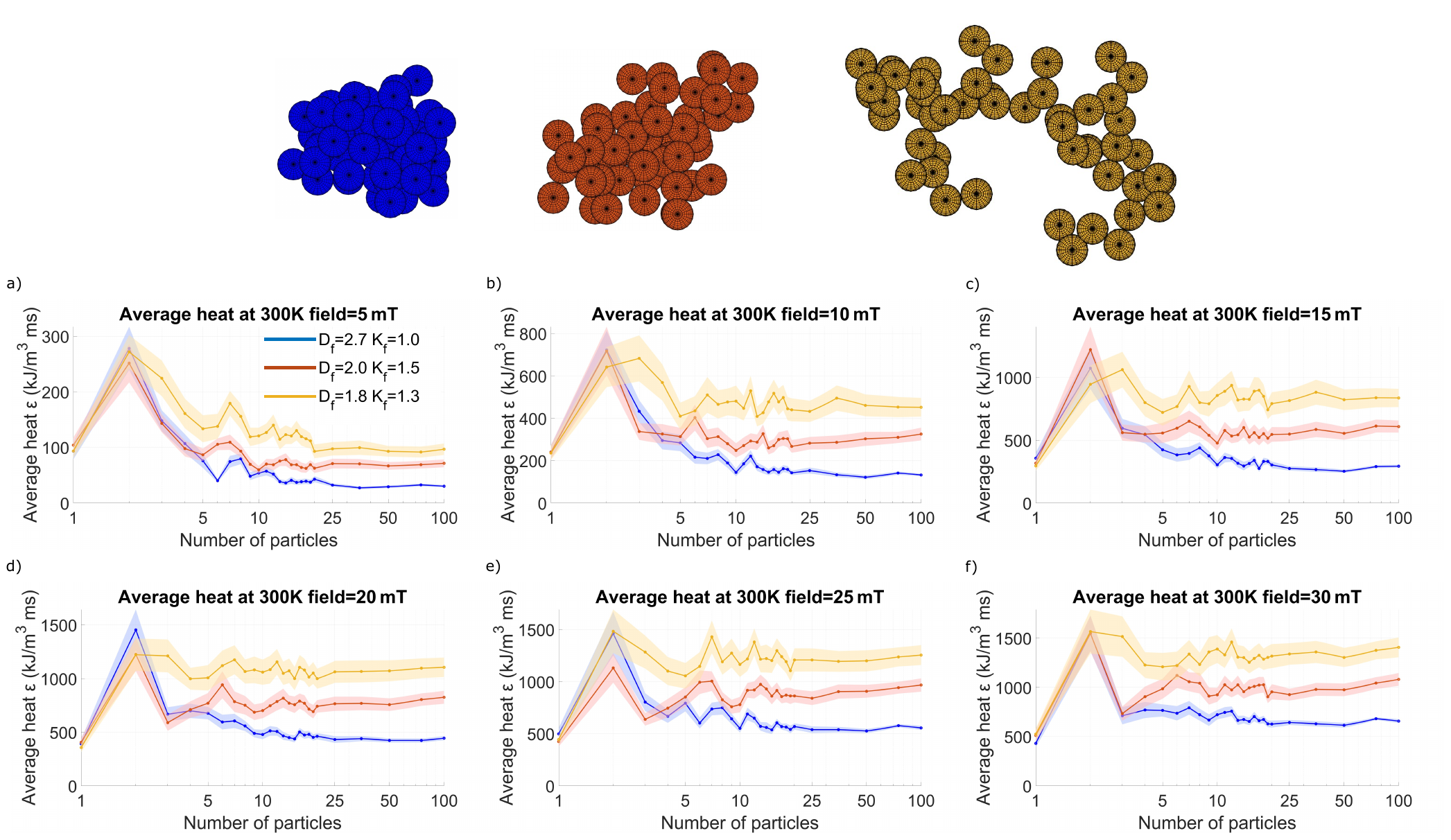}
  \caption{Average generated heat per particle, and the uncertainty on this value (shaded regions) released by the different color-coded clusters (colored lines) as a function of the number of particles at 300 K in the presence of a sinusoidally varying applied field with a frequency of 300 kHz and  amplitudes ranging from 5 mT to 30 mT. Three different sets of fractal parameters are shown: $D_\mathrm{f}=2.7$ $k_\mathrm{f}=1$ (blue); $D_\mathrm{f}=2$ $k_\mathrm{f}=1.5$ (red) and $D_\mathrm{f}=1.8$ $k_\mathrm{f}=1.3$ (yellow).}
  \label{fig:Heat300Kwitherrors.pdf}
\end{figure*}

By investigating the athermal case we were able to determine a clear picture of the energy landscapes, which will allow us to understand the heat generation in the presence of thermal switching, which is the topic of this section. The results of the simulations obtained at 300 K are shown in Figs. \ref{fig:Heat300Kwitherrors.pdf} and \ref{HystLoops.pdf} b).

For the particles under study, the N\'eel relaxation time is estimated at about 1 $\mu$s, which means that we expect substantial thermal switching on the timescale of our 300 kHz excitation field, and therefore a much lower heating than the one obtained at 0 K. Indeed, whereas the single particle heating saturated at about 3000  kJ m$^{-3}$ms$^{-1}$ at 0 K, it already saturates at 500 kJ m$^{-3}$ms$^{-1}$ at 300 K, as also clearly visible in the single particles' hysteresis loop. Note that this effect is very pronounced for the particles under study here but strongly diminishes for larger particles, with longer N\'eel relaxation times. Nonetheless, in the presence of thermally assisted switching, the generated heat is almost always lower than in the athermal case, except when the excitation field strength was insufficient to overcome the switching barriers, i.e. was lower than the coercivity of the clusters. In the hysteresis loops, this is clearly visible as all loops display a lower coercivity at 300 K than at 0 K, and all loops close already around 20 mT, indicating that higher fields will not necessarily result in additional heating. This effect is confirmed in panels e) and f) where the heat release is similar for both cases.

Using the same picture of the energy barriers that we deduced in the athermal case, the larger energy barriers for the 2-particle clusters hamper the (premature) thermal switching, allowing the field to reach higher values before a switching event takes place, resulting in hysteresis loops with a coercive field that is almost twice as large as compared to the single particle case, and consequently results in an increased heat generation. Qualitatively, the shape of these images, and their underlying physics, is therefore more similar to the ones obtained at the highest field intensities in the athermal case.

The differences between the compact and more elongated clusters are qualitatively similar, again with elongated clusters generating more heat than the compact ones. However, because of the exponential dependence of the thermally assisted switching on the height of the energy barriers, the  fractal dimensions plays a larger role here. This is reflected in a larger difference in the shape of the hysteresis loops corresponding to the compact and elongated clusters, already visible for clusters containing as few as 3 particles. This effect is  most notably when the applied field has an intensity similar to the coercive field, $B_c=\frac{2K}{Ms}=22.4$ mT, as seen by comparing panels c) and d) in Figures \ref{fig:Heat0Kwitherrors.pdf} and \ref{fig:Heat300Kwitherrors.pdf}.

Additionally, it is important to notice that, again, the average released heat is approximately constant for aggregates containing more than 20 particles, reasserting the conclusions found in the previous section.  
%
%


\subsection{Average heat per particle as a function of fractal parameters}
\label{sec:parameters}

As we have just discussed, the average heat per particle stabilizes with respect to the cluster size, and from about 20 particles onward reaches a value that becomes independent of cluster size. Instead, it only depends on the specific configuration and orientation of each cluster, which allows to extrapolate these results to bigger clusters.
To investigate the dependence on the fractal parameters in detail, we simulated a larger set of parameters at 300 K using 100 realizations of the aggregate geometries, containing 50 particles to ensure a stable heating. Using the Matlab Curve Fitting Toolbox$^{TM}$\cite{matlabcurvefitting} we fitted an equation to find the average heat per particle as a function of the fractal parameters $D_\mathrm{f}$ and $k_\mathrm{f}$. 
We found that the heating $Q$ is described well by a plane, defined by linear dependencies on both 
$D_\mathrm{f}$ and $k_\mathrm{f}$: 
  \begin{equation}
      Q(D_f,k_f) = A D_f+ B k_f+ C
  \end{equation}
  with prefactors $A=- 118.5 \pm18.6 \,$ kJ m$^{-3}$ms$^{-1}$, $B=- 54.52 \pm16.93 \,$ kJ m$^{-3}$ms$^{-1}$, and an offset $C$ of $4601 \pm 579\,$ kJ m$^{-3}$ms$^{-1}$.

\begin{figure}[ht]
\centering
  \includegraphics[width=8.6cm]{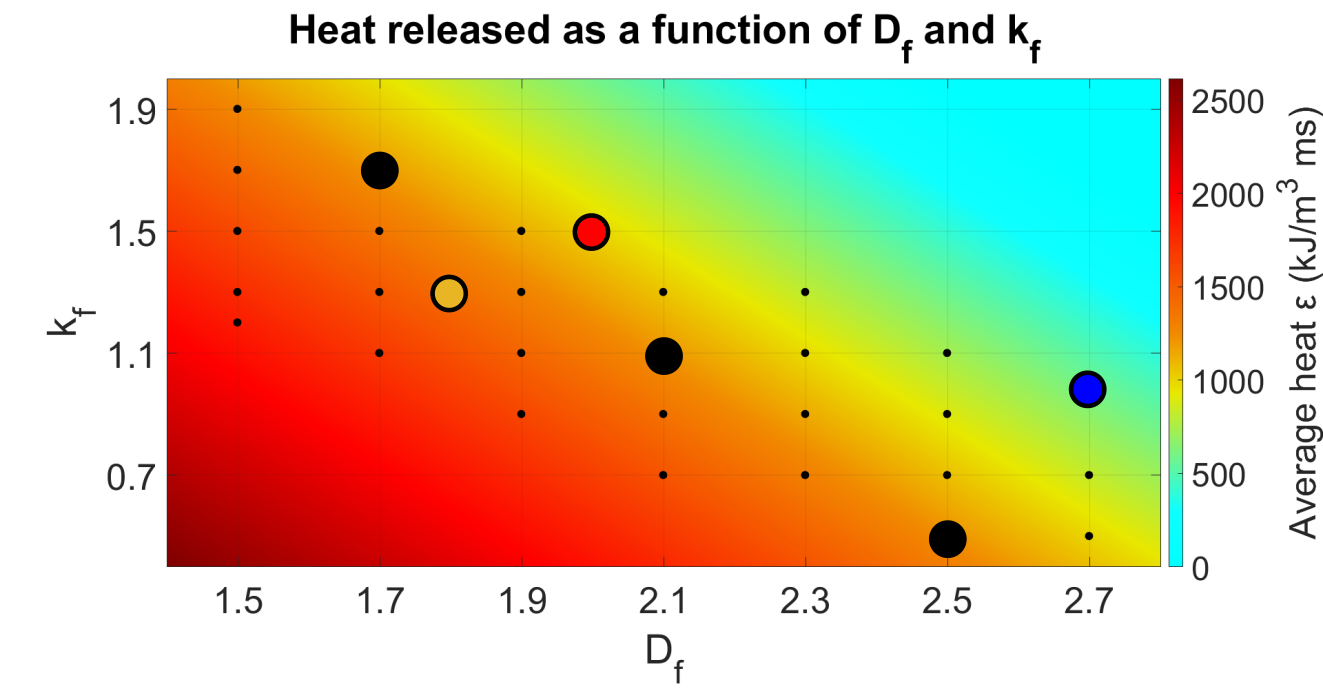}
  \caption{Average heat per particle as a function of the fractal dimensions \(D_{f}\) and \(k_{f}\) at 300 K. The three points highlighted in blue, red and yellow match the exemplary aggregates studied in the previous section, as shown on top of Figures \ref{fig:Heat0Kwitherrors.pdf} and \ref{fig:Heat300Kwitherrors.pdf}. Highlighted black points represents aggregates that dissipate similar amounts of heat while having different parameters, which can visualized in Table \ref{tab:equalheat}.}
  \label{fig:HeatVsPara.png}
\end{figure}

The results of the simulations and the fit are presented in Figure \ref{fig:HeatVsPara.png}. The data points highlighted in blue, red and yellow, correspond to the clusters depicted in both Figs. \ref{fig:Heat0Kwitherrors.pdf}  and \ref{fig:Heat300Kwitherrors.pdf}, with parameters extracted from clusters found in biological samples \cite{Etheridge2014}. This figure further confirms that the cluster shape strongly affects the heating, being driven by both the fractal dimension and the fractal prefactor, which together define the fractal shape. Both values are inversely related to the heating, in agreement with literature \cite{Wozniak2012}. Additionally, different sets of parameters can give rise to similarly looking clusters, which also release similar amounts of heat.
As an example, the three data points highlighted in black are displayed in more detail in Table \ref{tab:equalheat}, and all display the same heating.

\begin{table}[]
    \centering
    \begin{tabular}{|c|c|c|c|}
    \hline
    
         $k_f$&1.7&1.1&0.5  \\
         \hline
         $D_f$&1.7 &2.1&2.5\\
         \hline
        Cluster&\includegraphics[width=2cm]{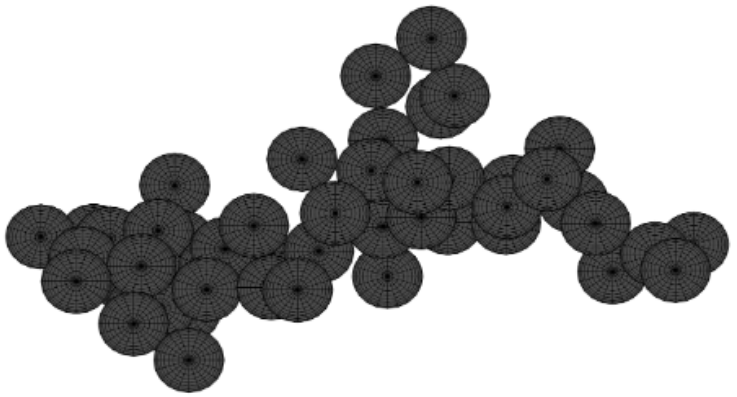}&
         \includegraphics[width=2cm]{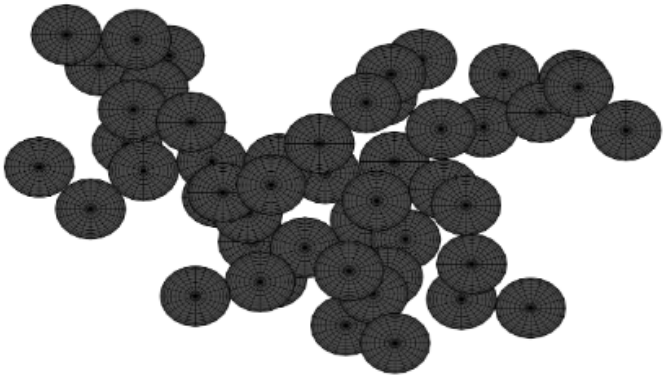}
         &\includegraphics[width=2cm]{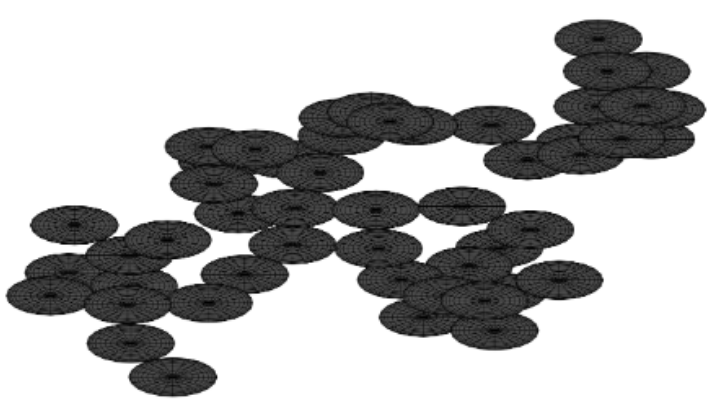} \\ 
    \hline
    \end{tabular}
    \caption{Example of 3 different clusters with different fractal parameters $k_f$ and $D_f$, which result in visually similar clusters and almost identical heat generation.}
    \label{tab:equalheat}
\end{table}
From this observation we conclude that when extracting the fractal parameters from experimental data, $D_f$ and $k_f$ are not independent parameters. From a practical point of view, an entire range of good fits can be made, thus allowing to fix the value of $k_f$ and only fitting the fractal dimension. A common value for $k_f$ is 1.593, corresponding to an infinitely large hexagonal close packed aggregate of monodisperse hard spheres in contact\cite{Wozniak2012}.

\section{Conclusions}

In this manuscript, we investigated the average heat per nanoparticle released by clusters of nanoparticles described by a fractal structure, as they appear in biological samples. 
These clusters were subject to a temperature of 300 K and were excited by a sinusoidally varying field. We investigated the dependence of the released heat as a function of the size and the fractal parameters of the clusters.

Our main conclusions are that already starting from relatively small clusters of about 20 particles, the heat released per particle stagnates. This result proves that the heat emitted by large clusters can be predicted by simulating moderately sized clusters with the same geometrical and magnetic parameters.
The most important application of our findings is in magnetic hyperthermia treatment planning, since one of the major problems currently faced is to estimate the heat produced by an indefinite aggregate of nanoparticles after injection.

Furthermore, we showed that the heat release is strongly affected by the shape of the clusters. Using the value of the heat released by single Stoner-Wohlfarth particles at 0 K as a reference, we found that for the parameters used in this study, the most elongated chain-like clusters display about half as much heating for large excitations whereas the most compact clusters display about 1/5th  of this value. Even though the exact numbers will depend on the details of the excitation amplitude and frequency, particle size (distribution), and material composition, these values can be used as a rule of thumb.

\section*{Appendix}
Because our model considers single-domain particles, we can use the Stoner-Wohlfarth model\cite{stoner1948} to estimate the expected heating for ``clusters'' containing only a single particle.

\begin{figure}[ht]
\centering
  \includegraphics[width=8.6cm]{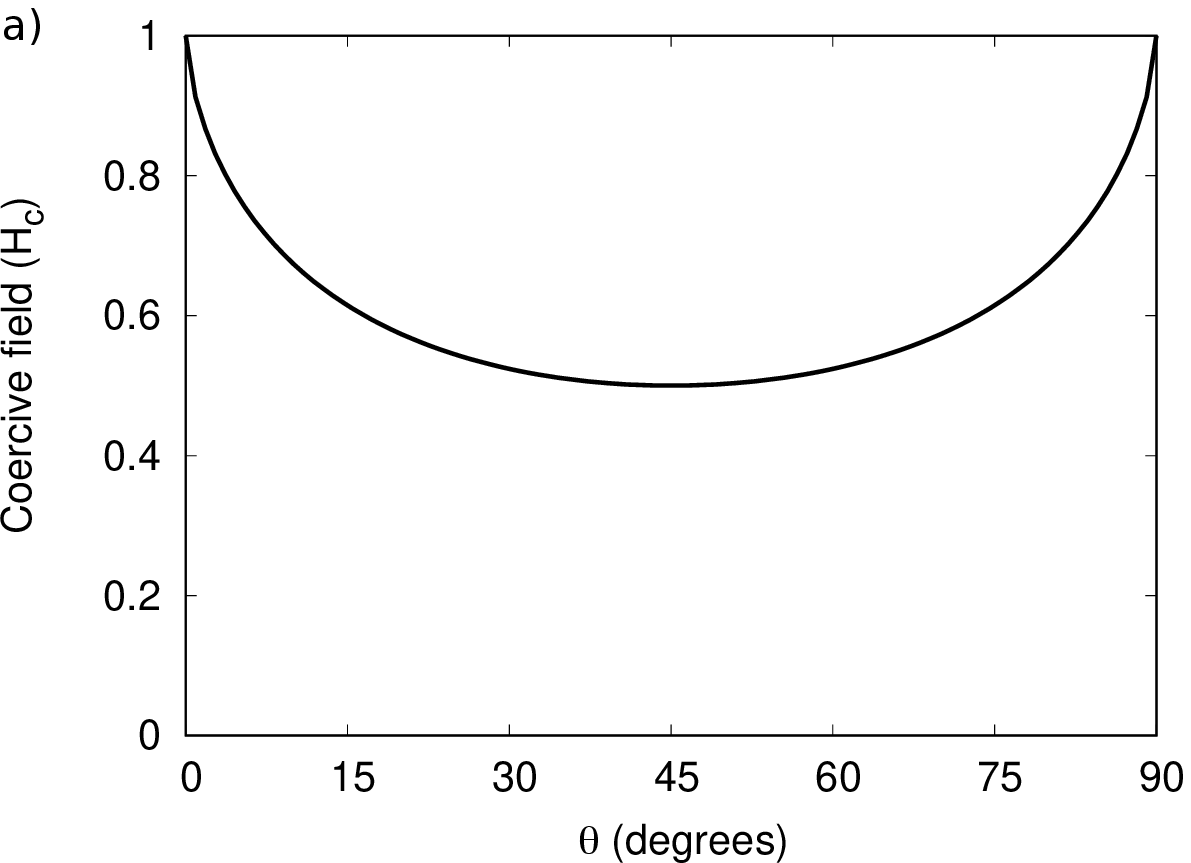}
  \includegraphics[width=8.6cm]{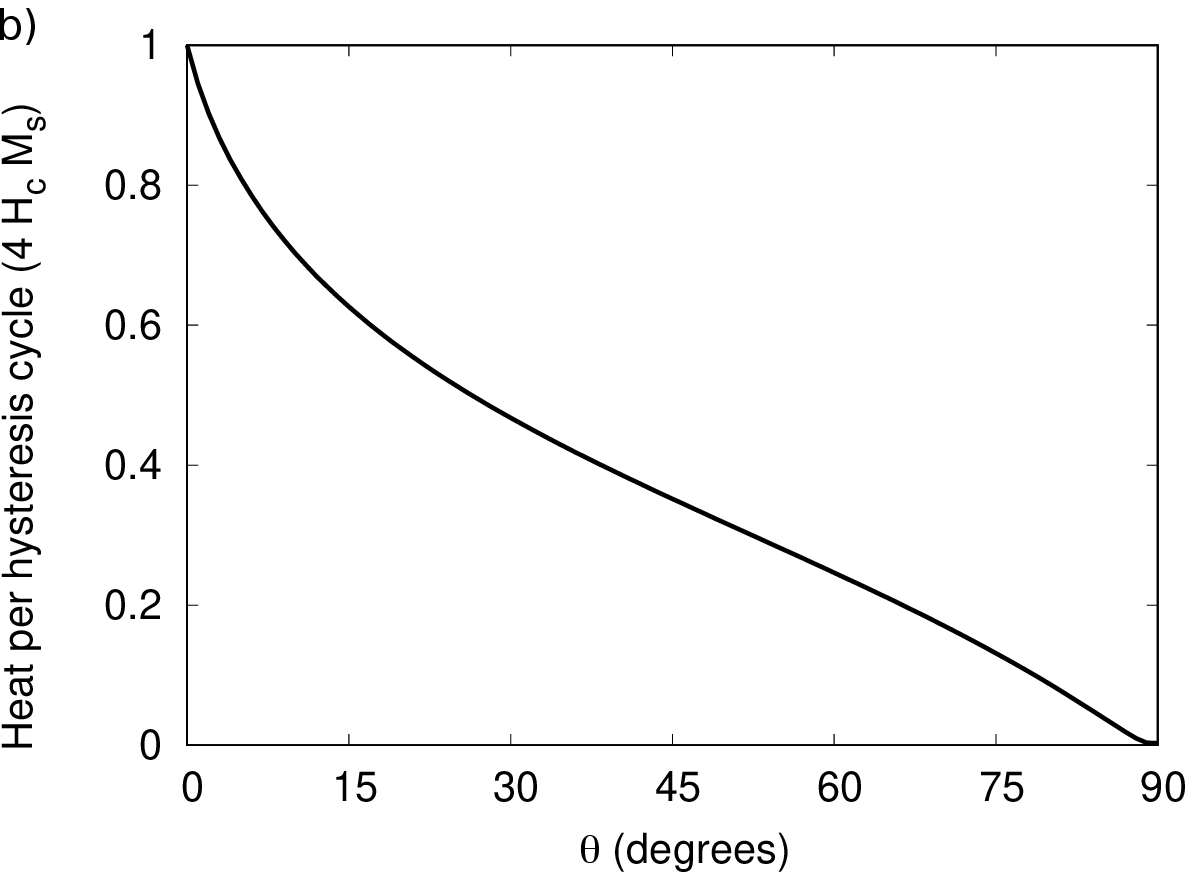}
  \caption{a) The coercive field of a Stoner-Wohlfarth particle as function of the angle $\theta$ between the applied field and the anisotropy axis. b) The heat dissipated in a single hysteresis cycle as function of $\theta$ of a single Stoner-Wohlfarth particle.}
  \label{fig:stoner}
\end{figure}

The coercive field that needs to be overcome for a particle to switch depends on the angle $\theta$ between the anisotropy axis and the externally applied field, and reaches a maximum of $H_c=\frac{2K}{M_s}$ for $\theta=0^{\circ}$  (i.e. with the field parallel to anisotropy axis), and a minimum of $H_c/2$ for $\theta=45^{\circ}$, as shown in Fig. \ref{fig:stoner} a). 

The heat dissipated in a single hysteresis cycle strongly depends on this angle as well, and quickly reduces to half of its maximum value of 4 $H_c M_s$ for $\theta$ as small as $25^{\circ}$, as shown in Fig. \ref{fig:stoner} b). As we consider ensembles with a anisotropy direction that is randomly, but uniformly, distributed on the unit sphere, large $\theta$ are more prevalent than small ones (with a proportionality $\propto \sin(\theta)$). This results in an average heat per hysteresis cycle that is only about 1/4th of the maximum value, i.e. $\sim H_c M_s$ (and a variation that is about equally large).

When considering the heating as function of field amplitude, we again turn our attention to Fig. \ref{fig:stoner} a), which shows the coercive field as a function of the angle between the anisotropy and the externally applied field, as given by 
\begin{equation}
    H_c=\frac{\sqrt{1-t^2+t^4}}{1+t^2}
\end{equation} with $t=\tan^{1/3}(\theta)$, as analytically determined by \citet{stoner1948}.

From this figure, we learn that, as soon as the external field is a bit larger than $H_c/2$, already a broad range of particle (with angles distributed symmetrically around $45^{\circ} $ can switch, while the external field needs to reach the full value of $H_c$ for \emph{all} particles to switch. However, as there are very few particles with $\theta=0^{\circ} $, and the ones with $\theta=0^{\circ}$ do not contribute to the heating at all, the average heat as function of field amplitude reaches its maximum value for quite soon for field strengths above $H_c/2$ and only slightly rises further for fields equal to or larger than $H_c$.  This is confirmed in section \ref{sec:0K}, in which $H_c/2\approx12$ mT, and the total heating observed for one particle clusters is zero for fields of 0 and 5 mT, and remains about constant for fields larger than 15 mT.

\section*{Conflicts of interest}
The authors declare no conflicts of interest.

\section*{Acknowledgements}
J.L. was supported by the Fonds Wetenschappelijk Onderzoek (FWO-Vlaanderen) with postdoctoral fellowship No. 12W7622N.
Part of the computational resources (Stevin Supercomputer Infrastructure) and services used in this work were provided by the VSC (Flemish Supercomputer Center), funded by Ghent University, FWO and the Flemish Government – department EWI.
This work has been also supported by the NoCanTher project, which has received funding from the European Union’s Horizon 2020 research and innovation programme under grant agreement no 685795. The authors acknowledge support from the COST Association through the COST action “MyWAVE” (CA17115). D. O. and J. O. J. are thankful for funding support from the Spanish Ministry of Science, Innovation under Contract no. PEJ2018-004866-A, grant PID2020-117544RB-I00, the Ramón y Cajal grant RYC2018- 025253-I and Research Networks grants RED2018-102626-T. The support from the Ministry of Economy and Competitiveness through the grants MAT2017-85617-R and the “Severo Ochoa” Program for Centers of Excellence in R\&D (SEV-2016-0686) is acknowledged. We would also like to thank the Systems Unit of the Information Systems Area of the University of Cádiz for computer resources and technical support.




\bibliography{rsc} 
\bibliographystyle{rsc} 

\end{document}